\documentclass[nofootinbib,prd,twocolumn,showpacs,showkeys,preprintnumbers]{revtex4-1}
\usepackage{hyperref,amssymb,amsmath,mathrsfs,bm,graphicx}
\usepackage[dvipsnames]{xcolor}
\begin{document}

\title {Braneworld-Klein-Gordon system in the framework of gravitational decoupling}

\author{P. Le\'on}
\email{pablo.leon@ua.cl}
\affiliation{Departamento de F\'isica, Universidad de Antofagasta, Aptdo 02800, Chile.}

\author{A. Sotomayor}
\email{adrian.sotomayor@uantof.cl}
\affiliation{Departamento de Matem\'aticas, Universidad de Antofagasta, Aptdo 02800, Chile.}

\begin{abstract}
We analyze the effective field equations of the Randall-Sundrum braneworld coupled with a Klein-Gordon scalar field through the minimal geometric deformation decoupling method (MGD-decoupling). We introduce two different ways to apply the  MGD-decoupling method to obtain new solutions for this enlarged system. We also compare the behavior of the new solutions with those obtained directly from the Randall-Sumdrum braneworld without coupling to the scalar field.
\end{abstract}

\keywords{BraneWorld, Gravitational Decoupling, Scalar field, Black Holes}

\maketitle

\section{Introduction}
The theory of General Relativity (GR) is both the simplest and the most successful theory for describing the gravitational interaction. Over the last years plenty of experimental data that match predictions of GR have been reported. However, despite the great success of GR, the theory does not provide satisfactory explanations to some phenomena, such as dark matter and dark energy problems. This is one of the reasons that have lead physicists to consider theories beyond GR that could explain some of these problems. These theories may include different potentials in terms of scalar fields to model dark energy and dark matter (see \cite{Sotiriou} for a brief review of generalized scalar-tensor gravity theories) which can be considered either a fundamental aspect of the theory or an effective description of a more fundamental theory. Besides their cosmological applications, the scalar fields can have a particular interest in the study of the no-hair conjecture (see \cite{Hawking:2016msc,thomas1,babi,thomas2,radu,marcelobh,thomas3,kanti1,adolfo1,adolfo2,adolfo3,galtsov1,kanti2,kanti3,galtsov2,mtz,konstantin,Sotiriou2}).  

In addition to dark matter and dark energy problems, there is also a fundamental problem in physics known as the hierarchy problem, which consists of the huge scale difference between the gravitational and the weak interactions. In \cite{Randall1,Randall2}, Randall and Sundrum proposed a theory known as Randall-Sundrum Braneworld (RSBW) which can explain this problem. In the simplest models of RSBW, all the gauge interactions described by the standard model of particles with all our observable universe are confined to live in a 3-brane, embedded in a five dimensional space-time called the bulk. In contrast with gauge interaction, the gravitational one is not restricted to live in the 3-brane and can spread into the bulk. This means that in our observable universe (the 3-brane) we can only see a fraction of the gravitational interaction, which explain why the gravitational interaction is so weak compared to the Planck scale. 

Currently, there is not direct experimental data that could support the RSBW theory \cite{Aad:2012tfa,CMS:2012yf}. However, the study of the Einstein's field equations that includes the contribution from the RSBW model is justified and interesting, since this theory can explain the hierarchy problem.  Furthermore, it should be mentioned that a good agreement between the RSBW predictions and the dark matter observations was reported in \cite{Gergely:2011df} (see also \cite{Visinelli:2017bny,Vagnozzi:2019apd}). Now,  it is also important to mention that are other realizations of RSBW in higher dimensions (see for example \cite{mpg}). In the context of RSBW, there are also some cosmological models which consider the existence of an scalar field to study perturbation of the RSBW scenario and could provide some information about dark matter and dark energy problems.

Now, even when RSBW based models are very promising and a covariant formulation of the theory in five dimensions is known, there are some open problems associated with these models, mainly due to  the lack of solutions associated with the equations of motion in the complete five dimensional theory \cite{Germani,Kanti,Abdolrahimi,Dai,Nakas:2020sey}. One convenient way to clarify these problems and give some information about the impact of RSBW on gravity is based on the study of the effective field equations in four dimensions (our observable universe). In this case the contribution which arises from the five dimensional model can be interpreted as a contribution to the energy momentum tensor in the four dimensional GR equations.

In recent years, a new method was proposed, which is known as the minimal geometric deformation decoupling method (MGD-decoupling) and among his many applications allows to find new physically relevant solutions of Einstein’s equations through a very simple and elegant way. This method was originally developed in RSBW  \cite{Ovalle1,Ovalle15,Ovalle2,Ovalle16,Ovalle8,Ovalle9,Ovalle7,Ovalle10,Ovalle11} and later on used in General Relativity \cite{Ovalle}. 

Specifically, the method allows to study gravitational systems whose modified Einstein-Hilbert action is given by 

\begin{eqnarray}
S = \int \left[\frac{R}{2k^2}+\mathcal{L}\right]\sqrt{-g}d^4x + \alpha \mbox{(correction)}. 
\end{eqnarray}
Thus, the energy momentum tensor is given by
\begin{equation}
    \hat{T}_{\mu \nu} = T_{\mu \nu} +  \alpha \theta_{\mu \nu},
\end{equation}
where the source $\theta_{\mu \nu}$ could have different interpretations, for example another matter fluid (see for example \cite{Our,Estrada1,Gabbanelli,Morales1,Morales2,Tello2,Contreras7}), the coupling of  Einstein's equations with other fields like a Klein-Gordon scalar field (see \cite{Ovalle13}), the corrections coming from gravitational theories beyond GR (for example \cite{Sharif3,Sharif4,Sharif5,Tello4,Estrada3,Leon}). 

Originally, the method was restricted to spatial metric deformations in spherically symmetric systems. However, in the last years the method has been extended to include the formulation in alternative coordinates \cite{Our2}, spatial and temporal deformations of the metric components \cite{Ovalle12} and cylindrical or axially symmetric matter distributions \cite{Sharif7,Contreras13} (for more applications of the gravitational decoupling method see \cite{Abellan,Abellan3,Sharif,Sharif2,Sharif8,Sharif9,Ovalle6,Ovalle17,Ovalle18,Ovalle19,Cavalcanti,Darocha1,Darocha2,Darocha3,Darocha4,Darocha5,Darocha6,Darocha7,Darocha8,Darocha9,Casadio2,Contreras,Contreras5,Contreras14,Contreras15,Rincon,Rincon2,Tello6,Tello7,Hensh,Maurya2,Maurya3,Maurya4,Maurya5,Zubair} ). 

In this paper, we study the RSBW effective GR equations coupled to a Klein-Gordon (KG) scalar field using the MGD-decoupling method. Is this formulation, the coupling to the scalar field on the observable brane is interpreted as the effective contribution of the coupling in the bulk (see \cite{Brax}). The  energy-momentum tensor of effective field equations can be decomposed into two sources, one associated with the corrections of the RSBW, and the other with the KG scalar field. In the context of GR, the order in which the MGD-decoupling is applied to decouple the different sources $\theta_i^{\mu \nu} $ is not relevant, but is not so clear when the sources are contributions from theories beyond GR. In fact, we will show that the MGD method leaves us with two inequivalent possibilities studying this enlarged system, depending on the the order in which the decoupling of the source is made.

The paper is organized as follows: in section 2 we summarize the MGD method. In section 3  we  present the  effective RSBW four dimensional Einstein's equations and in section 4 we introduce two ways in which the MGD method can be applied to analyze RSBW model coupled with a Klein Gordon scalar field. In section 5 we give our conclusions.

\label{Introdcution}

\section{Gravitational decoupling method}

\label{The MGD method}
\label{section MGD}
The so-called gravitational decoupling method (MGD-decoupling) \cite{Ovalle} had its motivation in the context of minimal geometric deformation (MGD), applied to the brane world model \cite{Randall1,Randall2,Ovalle20} (to read an elegant and detailed exposition relating these aspects see \cite{Ovalle15}) . 

To give a concise review of the method, we start with Einstein field equations (EFEs)

\begin{equation}
\label{2.1}
R_{\mu\nu}-\frac{1}{2}\,R\, g_{\mu\nu}=-k^2\hat{T}_{\mu\nu},
\end{equation}
where we assume that the energy momentum tensor $\hat{T}_{\mu \nu}$ has contributions  of two (different) gravitational sources given by

\begin{equation}
\hat{T}_{\mu \nu}=T_{\mu \nu} + \Theta_{\mu \nu}.
\label{2.2}
\end{equation}

In this work, we are concerned with the case of spherically symmetric and static systems. We use the well-known Schwarzschild-like coordinates with the corresponding line element given by

\begin{equation}
ds^{2}=e^{\nu (r)}\,dt^{2}-e^{\lambda(r)}\,dr^{2}-r^{2}\left( d\theta^{2}+\sin ^{2}\theta \,d\phi ^{2}\right)
\label{2.3}.
\end{equation}

In this setting, Einstein's field equations (\ref{2.1}) can be written in the known form

\begin{eqnarray}
\label{2.4}
k^2 \bar{\rho} = k^2 (T^0_0 + \Theta_0^0)  &=&\frac{1}{r^2}-e^{-\lambda}\left(\frac{1}{r^2}-\frac{\lambda'}{r}\right)\ ,\\
\label{2.5}
k^2 \bar{p}_r = - k^2 (T^1_1 + \Theta_1^1) &=&-\frac 1{r^2}+e^{-\lambda}\left( \frac 1{r^2}+\frac{\nu'}r\right)\ ,\\
\label{2.6}
k^2 \bar{p}_t = -k^2 (T^2_2 + \Theta_2^2) &=&\frac{e^{-\lambda}}{4}\Bigg( 2\nu''+\nu'^2- \lambda' \nu' \nonumber \\ & + & \left. 2\frac{\nu'-\lambda'}{r}\right).
\end{eqnarray}
The prime indicates a derivative with respect to $r$, and $\bar{\rho}$, $\bar{p}_r$ and $\bar{p}_t$ are defined to be the effective energy density, the effective radial pressure and the effective tangential pressure, respectively.

The conservation equation for the given system is written as

\begin{equation}
\nabla_\mu \hat{T}^{\mu \nu} = (\bar{p}_r)'-\frac{\nu'}{2}(\bar{\rho}+\bar{p}_r)-\frac{2}{r}(\bar{p}_t-\bar{p}_r) =0
\label{2.8}
\end{equation}
which, in terms of the gravitational sources $T_{\mu \nu}$ and $\Theta_{\mu \nu}$, takes the following form
\begin{eqnarray}
\label{2.9}
\ & & \left({T}_1^{\ 1}\right)'-\frac{\nu'}{2}\left({T}_0^{\ 0}-{T}_1^{\ 1}\right)-
\frac{2}{r}\left({T}_2^{\ 2}-{T}_1^{\ 1}\right)
\nonumber \\ &+& \left({\Theta}_1^{\ 1}\right)'-\frac{\nu'}{2}\left({\Theta}_0^{\ 0}-{\Theta}_1^{\ 1}\right)-\frac{2}{r}\left({\Theta}_2^{\ 2}-{\Theta}_1^{\ 1}\right)
=0.
\end{eqnarray}

We notice, from equations (\ref{2.4})-(\ref{2.6}), that the combination of the two sources in the energy-momentum tensor describes a fluid with local anisotropy on the pressures.

In order to solve the complete system of equations (\ref{2.4})-(\ref{2.6}), we apply the MGD-decoupling method. The first step considers only the contribution of the source $T_{\mu \nu}$, whose line element is  written 

\begin{equation}
ds^{2}=e^{\xi (r)}\,dt^{2}-\frac{1}{\mu(r)}\,dr^{2}-r^{2}\left( d\theta^{2}+\sin ^{2}\theta \,d\phi ^{2}\right),
\label{2.10}
\end{equation}
where

\begin{equation}
\label{2.11}
\mu(r)\equiv 1-\frac{k^2}{r}\int_0^r x^2\,T^0_0\, dx
=1-\frac{2\,m(r)}{r},
\end{equation}
is the standard definition of the mass function in General Relativity (GR).

The next step also includes the contribution of  $\Theta_{\mu \nu}$. The effects induced by this gravitational source are encoded by deformations of temporal and radial components in the metric, given by

\begin{eqnarray}
\label{2.12}
\nu(r) & = & \xi(r)+ \alpha g^*(r), \\
\label{3.5}
e^{-\lambda(r)}  & = & \mu(r)+ \alpha f^*(r) ,
\end{eqnarray}
where $g^*(r)$ and $f^*(r)$ are two functions to be determined. 

In this work, we are only concerned with the case where $g^*=0$ (known as minimal geometric deformation), that is, when we have only deformations of the radial component of the metric in the reduced form 

\begin{eqnarray}
\label{2.13}
\nu(r) & = & \xi(r), \\
e^{-\lambda(r)}  & = & \mu(r)+ \alpha f^*(r)\label{14}.
\end{eqnarray}
This transformation has been used in the context of RSBW in \cite{Ovalle1,Ovalle15,Ovalle2,Ovalle16,Ovalle8,Ovalle9,Ovalle7,Ovalle10,Ovalle11,Leon}. Now, by using (\ref{2.13}) and (\ref{14}), Einstein's field equations can be decoupled into two systems of equations. 

The first one is given by Einstein field equations relative to the source $T_{\mu \nu}$ 

\begin{eqnarray}
\label{2.14}
k^2 T_0^0 & = & \frac{1}{r^2} -\frac{\mu}{r^2} -\frac{\mu'}{r}\ , \\
\label{2.15}
-k^2 T_1^1 & = & -\frac 1{r^2}+\mu\left( \frac 1{r^2}+\frac{\nu'}r\right)\ , \\
\label{2.16}
-k^2 T_2^2 & = & \frac{\mu}{4}\left(2\nu''+\nu'^2+\frac{2\nu'}{r}\right)+\frac{\mu'}{4}\left(\nu'+\frac{2}{r}\right) \ ,
\end{eqnarray}
with associated conservation equation
\begin{eqnarray}
\label{2.17}
\left({T}_1^{\ 1}\right)'-\frac{\nu'}{2}\left({T}_0^{\ 0}-{T}_1^{\ 1}\right)-
\frac{2}{r}\left({T}_2^{\ 2}-{T}_1^{\ 1}\right) = 0,
\end{eqnarray}
while the second system is only related to the gravitational source $\Theta_{\mu \nu}$, and written by

\begin{eqnarray}
\label{2.18}
k^2 \Theta_0^{\,0}
&\!\!=\!\!&
-\frac{\alpha f^{*}}{r^2}
-\frac{\alpha f^{*'}}{r}
\ ,
\\
\label{2.19}
k^2 \Theta_1^{\,1}
&\!\!=\!\!&
-\alpha f^{*}\left(\frac{1}{r^2}+\frac{\nu'}{r}\right)
\ ,
\\
\label{2.20}
k^2 \Theta_2^{\,2}
&\!\!=\!\!&
-\alpha \frac{f^{*}}{4}\left(2\nu''+\nu'^2+2\frac{\nu'}{r}\right)
-\alpha \frac{f^{*'}}{4}\left(\nu'+\frac{2}{r}\right)
\ ,
\end{eqnarray}
with corresponding conservation equation 
\begin{eqnarray}
\label{2.21}
\left(\Theta_1^{\,\,1}\right)'
-\frac{\nu'}{2}\left(\Theta_0^{\,\,0}-\Theta_1^{\,\,1}\right)
-\frac{2}{r}\left(\Theta_2^{\,\,2}-\Theta_1^{\,\,1}\right)
=
0.
\end{eqnarray}

To convert the system (\ref{2.18})-(\ref{2.20}) into a Einstein system for $\Theta_{\mu \nu}$, we can redefine the components of $ \Theta_{\mu \nu} $ in order to include the corresponding lost factors of $ 1/r^2 $. Equations (\ref{2.9}), (\ref{2.19}) and (\ref{2.20}) show that interaction between the sources $T_{\mu \nu}$ and $\Theta_{\mu \nu}$ is purely gravitational. 

Thus, to decouple the complete Einstein system (\ref{2.4})-(\ref{2.6}), we first proceed by solving the Einstein system (\ref{2.14})-(\ref{2.17}) for the original source $T_{\mu \nu}$, determining the triplet $\{T_{\mu \nu},\xi,\mu\}$. The second step is to solve the system (\ref{2.18})-(\ref{2.20}) for the source $\Theta_{\mu \nu}$, to find $\{\Theta_{\mu \nu},f^*\}$. Finally, the solution for the complete system can be obtained through a combination of the results obtained in the previous two steps.    

This simple and systematic procedure, known as MGD-decoupling, can be used as a powerful tool in the analysis of more complicated and realistic distributions of matter in the context of General Relativity.

\section{The Braneworld context}
\label{The Braneworld coupled to a scalar field}

One of the main features of the braneworld models is that the five dimensional gravity induces modifications in the $(3+1)$ observable universe, called the brane, which can be expressed as 

\begin{equation}
G_{\mu \nu} = -g_{\mu \nu} \Lambda -k^2 \hat{T}_{\mu\nu},
\label{3.1}
\end{equation}
where $k^2=8\pi G_N$ and $\Lambda$ is the cosmological constant onto the brane.

The induced modifications to the Einstein field equations are given by the effective energy-momentum tensor

\begin{equation}
    \hat{T}_{\mu \nu}= T_{\mu \nu} - \frac{6}{\sigma}S_{\mu \nu} +\frac{1}{8\pi} \mathcal{E}_{\mu \nu} +\frac{4}{\sigma}\mathcal{F}_{\mu \nu},
    \label{3.2}
\end{equation}
which, through the inclusion of the last three terms, takes into account all the effects of the bulk onto the brane, with $\sigma$ being the brane tension.

The $S_{\mu \nu}$ high-energy term arises from the extrinsic curvature terms in the projected Einstein tensor onto the brane and is given by

\begin{equation}
    S_{\mu \nu}= \frac{1}{12}TT_{\mu \nu}-\frac{1}{4}T_{\mu \rho} T^\rho_\nu + \frac{g_{\mu \nu}}{24} \left[3T_{\rho \lambda} T^{\rho \lambda}-T^2\right], \label{3.3}
\end{equation}
where $T$ is the trace of $T_{\mu \nu}$. 

The $\mathcal{E}_{\mu \nu}$ Kaluza-Klein corrections term represents the projection of the Weyl tensor of the bulk. For the case of spherically  symmetric and static distributions of matter, which will be the case, it can by written as

\begin{equation}
    k^2 \mathcal{E}_{\mu \nu} = \frac{6}{\sigma}\left[\mathcal{U}\left(u_\mu u_\nu -\frac{1}{3}h_{\mu \nu}\right)+\mathcal{P}_{\mu \nu}\right], \label{3.4}
\end{equation}
with

\begin{eqnarray}
h_{\mu \nu} & = & g_{\mu \nu}-u_\mu u_\nu, \label{3.5} \\
\mathcal{P}_{\mu \nu} & = & \mathcal{P}\left(r_\mu r_\nu -\frac{1}{3}h_{\mu \nu}\right), \label{3.6}
\end{eqnarray}
where $\mathcal{U}$, $\mathcal{P}_{\mu \nu}$, $h_{\mu \nu}$, $u_\mu$ and $r_\mu$ are the bulk Weyl scalar, the anisotropic stress, the projection operator operator, the four velocity of the fluid element and the radial unitary vector, respectively.

The last correction to the effective energy-momentum tensor $\mathcal{F}_{\mu \nu}$ term depends on all the stresses but the cosmological constant in the bulk. From now on, we consider that $\mathcal{F}_{\mu \nu}=0$, which means that only the cosmological constant is present in the bulk. In this particular case, we recover the standard conservation equation of GR

\begin{equation}
    \nabla^\nu T_{\mu \nu}=0. \label{eqq}
\end{equation}

In order to study the effects of the RSBW on anisotropic fluids we use that the energy-momentum tensor $T_{\mu \nu}$ is given by

\begin{equation}
    T_{\mu \nu}=(\rho + p_t)u_\mu u_\nu -p_tg_{\mu \nu}+(p_r-p_t)s_{\mu}s_{\nu},
    \label{3.7}
\end{equation}
where $u_{\mu}=\exp{\nu/2}\delta^\mu_0$, $s_{\mu}=\exp{\lambda/2}$, and $\rho$,$p_r$,$p_t$ are the energy density and the radial and tangential pressures of the fluid, respectively. In this situation the equilibrium equation leads to

\begin{equation}
    p_r'+\frac{\nu'}{2}(\rho +p_r)-\frac{2\Delta}{r}=0, \quad \Delta=p_t-p_r. \label{eqpf}
\end{equation}

We are now ready to writte the effective Einstein's equations (when the cosmological constant $\Lambda=0$) for the four dimensional 3-brane. Using (\ref{2.10})  and (\ref{3.2})-(\ref{3.7}), equation (\ref{3.1}) leads to

\begin{eqnarray}
k^2 \Bigg[\rho  & + & \frac{1}{\sigma}\left(\frac{(\rho^2-\Delta^2)}{2}+\frac{6\mathcal{U}}{k^4}\right)\Bigg]  = \nonumber \\ && \frac{1}{r^2} -e^{-\lambda}\left(\frac{1}{r^2}-\frac{\lambda'}{r}\right) , \label{3.8} \\
k^2 \Bigg[p_r &+& \frac{1}{\sigma}\left(\frac{\rho^2}{2}+ \rho p_t + \frac{p_t^2-p^2_r}{2} +\frac{2\mathcal{U}}{k^4}\right) + \frac{4\mathcal{P}}{k^4\sigma}\Bigg]  = \nonumber \\ &-&  \frac 1{r^2}+ e^{-\lambda}\left( \frac 1{r^2}+\frac{\nu'}r\right),  \label{3.9} \\
k^2 \Bigg[p_t &+& \frac{1}{\sigma}\left(\frac{\rho^2}{2}+ \frac{\rho}{2}(p_r+p_t)+\frac{2\mathcal{U}}{k^4}\right) -\frac{2\mathcal{P}}{k^4\sigma}\Bigg]  = \nonumber \\ && \frac{e^{-\lambda}}{4}\left(2\nu''   +  \nu'^2- \lambda' \nu' +2\frac{\nu'-\lambda'}{r}\right) \label{3.10}.
\end{eqnarray}

We see from equations (\ref{3.8})-(\ref{3.10}), that we have reached to an indefinite system, in which extra information is required (related with the geometry of the bulk) in order to be solved.

In next section we show how (through MGD-decoupling), starting from given solutions of RSBW system, we can obtain solutions of the Braneworld-Klein-Gordon system. Also we show the analogue of the above procedure starting with solutions of the Einstein-Klein-Gordon system (see figure 1 in next page).

\section{RSBW coupled to a KG scalar field}
In this section we analyze the 4D RSBW effective equations minimally coupled with a KG scalar field using the MGD-method. In particular, we only be interested in the analysis of the external region of a spherically symmetric distribution satisfying these equations. 

\begin{figure}
    \centering
    \resizebox{0.5\textwidth}{!}{%
    \includegraphics{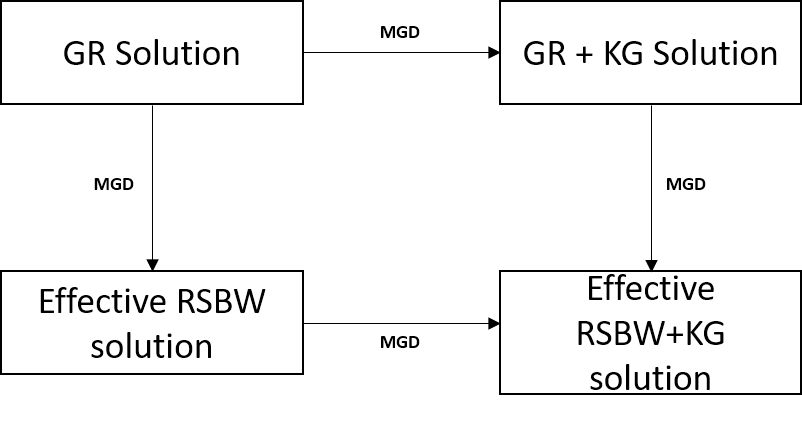}}
    \caption{Construction of the solution for the coupling of RSBW with KG}
    \label{Diagrama}
\end{figure}

The MGD method gives two options to study this enlarged system. First, we can choose a known solution for the effective field equations of the RSBW and use the method to get a solution for the complete system. The second approach is to select, as seed, a solution of the Einstein-Klein-Gordon equations and extend it to the RSBW scenario through the MGD method. Now, it is known that MGD method can been used to search solutions of the Einstein Klein Gordon and  RSBW independently (see \cite{Ovalle13,Leon}). In both cases, the seed is GR solution. This means that we can start with a GR solution and end up with a RSBW solution minimally coupled to a scalar field (see figure \ref{Diagrama}). 

We analyse in the following two possible ways to obtain solutions for the effective RSBW coupled to the Klein Gordon scalar field. 

In order to study the effect of a KG scalar field in the external solutions of the RSBW field equation we consider the following energy-momentum tensor  

\begin{equation}
T^T_{\mu \nu}= T_{\mu \nu}-\frac{6%
}{\sigma }\,S_{\mu \nu }+\frac{1}{8\pi }\,\mathcal{E}_{\mu \nu }
+\,\Theta_{\mu\nu}
\ , \label{tkg}
\end{equation}
where the first term is the energy momentum tensor of an anisotropic fluid (given by Eq. (\ref{3.7})), the following  two terms are the contributions of the BW sector and $\Theta_{\mu \nu}$ is a source coming from the presence of a KG scalar field filling the space time. 

Then $\Theta_{\mu \nu}$ is given by

\begin{equation}
    \Theta_{\mu \nu}=\nabla_\mu \Psi \nabla_\nu \Psi - \left(\frac{1}{2}\nabla_\alpha \Psi \nabla^\alpha \Psi -V(\Psi)\right)g_{\mu \nu}, \label{kg}
\end{equation}
where $\Psi(r)$ is a scalar minimally coupled and $V(\Psi)$ is a self interaction potential. 

Einstein's equations corresponding to the enlarged system (\ref{tkg}) can be written by

\begin{eqnarray}
k^2 \Bigg[\rho  & + & \frac{1}{\sigma}\left(\frac{(\rho^2-\Delta^2)}{2}+\frac{6\mathcal{U}}{k^4}\right) + \frac{1}{2}e^{-\lambda}\Psi'^2 + V \Bigg]  = \nonumber \\ && \frac{1}{r^2} -e^{-\lambda}\left(\frac{1}{r^2}-\frac{\lambda'}{r}\right) , \label{3.8} \\
k^2 \Bigg[p_r &+& \frac{1}{\sigma}\left(\frac{\rho^2}{2}+ \rho p_t + \frac{p_t^2-p^2_r}{2} +\frac{2\mathcal{U}}{k^4}\right)  +  \frac{4\mathcal{P}}{k^4\sigma} \nonumber \\ &+& \frac{1}{2}e^{-\lambda}\Psi'^2 - V\Bigg]  = -  \frac 1{r^2}+ e^{-\lambda}\left( \frac 1{r^2}+\frac{\nu'}r\right),  \label{3.9} \\
k^2 \Bigg[p_t &+& \frac{1}{\sigma}\left(\frac{\rho^2}{2}+ \frac{\rho}{2}(p_r+p_t)+\frac{2\mathcal{U}}{k^4}\right) \nonumber \\ &-&\frac{2\mathcal{P}}{k^4\sigma}- \frac{1}{2}e^{-\lambda}\Psi'^2 - V\Bigg]  =  \frac{e^{-\lambda}}{4}\left(2\nu'' \right. \nonumber \\ & + & \left. \nu'^2- \lambda' \nu' +2\frac{\nu'-\lambda'}{r}\right) \label{3.10}.
\end{eqnarray}

\subsection{First approach}

In this approach we choose, as seed, a solution of the 4D effective BRSW. Then, with the MGD-decoupling method, we couple it with the Klein-Gordon scalar field.

Following the procedure presented in section (\ref{section MGD}), the system of equations we need to solve is given by 

\begin{eqnarray}
k^2 \left(\frac{1}{2}e^{-\lambda}\Psi'^2 + V\right) & = & -\alpha \left(\frac{f^*}{r^2}+\frac{f^{*'}}{r} \right), \label{4.3} \\
k^2 \left(-\frac{1}{2}e^{-\lambda}\Psi'^2 + V\right) & = & - \alpha f^* \left(\frac{1}{r^2}+\frac{\xi'}{r}\right), \label{4.4} \\
k^2 \left(\frac{1}{2}e^{-\lambda}\Psi'^2 + V\right) & = & -\frac{\alpha f^*}{4}\left(2\xi'' + \xi'^2+2\frac{\xi'}{r}\right) \nonumber \\ &-& \frac{\alpha f^{*'}}{4}\left(\xi'+\frac{2}{r}\right) \label{4.5},
\end{eqnarray}
where the conservation equation can be written (using Eq. (\ref{2.21})) as

\begin{equation}
    \Psi''+\left(\frac{2}{r}+\frac{1}{2}(\xi'-\lambda')\right)\Psi' = e^{\lambda}\frac{dV}{d\Psi}.
\end{equation}

Combining these equations we can find the following differential equation for $f^*$:

\begin{equation}
    \left(\frac{\xi'}{4}-\frac{1}{2r}\right)f^{*'} + f^*\left(\frac{\xi''}{2}+\frac{(\xi^2)'}{4}+\frac{\xi'}{2r}-\frac{1}{r^2}\right)=0,
\end{equation}
whose solution is given by

\begin{equation}
    f^*= C\exp\left(\int \left(\frac{\xi''}{2}+\frac{\xi'^2}{4}+\frac{\xi'}{2r}-\frac{1}{r^2}\right)\Big{/}\left(\frac{1}{2r}-\frac{\xi'}{4}\right)dr\right). \label{fsl}
\end{equation}
Before give any particular example it will be useful to write $\Psi'^2$ as 

\begin{eqnarray}
\Psi'^2 & = & \frac{e^{\lambda}\alpha}{k^2 r}(f^{*}\xi'-f^{*'}), \label{psi} 
\end{eqnarray}
which can be re-expressed as
\begin{eqnarray}
\Psi_0-\Psi(r)=\pm\int_r^\infty e^{\lambda/2}\sqrt{\frac{\alpha}{k^2 \bar{r}}(f^{*}\xi'-f^{*'})}d\bar{r}, \label{cs}
\end{eqnarray}
where $\Psi_0 = \lim_{r\rightarrow 0} \Psi(r)$. In the same way, it is easy to get the expression for $V$ in terms of $f^*$·
\begin{eqnarray}
V & = & -\frac{\alpha}{2k^2}\left(f^*\left(\frac{2}{r^2}+\frac{\xi'}{r}\right)+\frac{f^{*'}}{r}\right). \label{pot}
\end{eqnarray}

We discuss now several examples of exterior solutions (with $\rho=p_r=p_t=0$), which will play the role of seed solutions for the MGD method. 
\subsubsection{First example}
First, we choose the following metric (see \cite{Ovalle11,Leon,Germani})

\begin{equation}
e^\xi =1-\frac{2M}{r}, \quad \mu=\left(1-\frac{2\,M}{r}\right)\left[1+\frac{D}{ \sigma \left(2r-3M\right)}\right]\ , \label{nss}
\end{equation}
where $D$ and $M$ are constants. 

Thus, from Eq. (\ref{fsl}) we obtain 

\begin{equation}
    f^*(r) = \left(1-\frac{2M}{r}\right)\left(\frac{C}{r-3M}\right)^2,
\end{equation}
and therefore
 
 \begin{eqnarray}
\Psi'^2 & = & \frac{2\alpha C^2}{k^2 r(r-3M)^3}\left[1+\frac{D}{ \sigma \left(2r-3M\right)} \right. \nonumber \\ &+& \left. \alpha\left(\frac{C}{r-3M}\right)^2\right]^{-1},  \nonumber \\ && \label{psip} \\
V & = & \frac{\alpha C^2 M}{k^2  r^2 (r-3 M)^3}. \label{pot}
 \end{eqnarray}

The $4D$ scalar curvature of the metric is given by

\begin{equation}
    R= -\frac{2\alpha C^2}{r(r-3M)^3}.
\end{equation}
This solution exhibits a naked singularity at $r=3M$. Therefore, this can not be interpreted as a black hole solution. However, if $r$ is always bigger than $3M$ this will represent a well behave external solution.

Now, Eq. (\ref{cs}) does not have an analytical solution for $\Psi$, at least not to all orders in $\alpha$ and $\sigma$. Thus, we shall perform a series expansion keeping terms up to first order in $\alpha$ and $\sigma^{-1}$. Therefore, we can integrate Eq. (\ref{cs}) to get
\begin{eqnarray*}
\Psi^2 & \approx &  \frac{8 C^2 \alpha r}{9k^2 M^2 (r-3 M)}\left(1-\frac{D}{3\sigma M}\right) \nonumber \\ &+& \frac{16C^2\alpha D}{27k^2\sigma M^3}\sqrt{\frac{r}{r-3M}}\arctan\left(\sqrt{\frac{r}{r-3M}}\right), \label{psi}
\end{eqnarray*}
where, for simplicity we choose
\begin{eqnarray}
\Psi_0 = -\frac{\sqrt{\alpha}|C|}{k}  \frac{  ((\pi -2) a R+12 M)}{9 \sqrt{2} M^2}.
\end{eqnarray}

The next step should be to write the scalar potential, in Eq. (\ref{pot}), as a function of $\Psi$. In order to do this, we shall use the Eq.(\ref{psi}) to find r as a function of $\Psi$ and introduce it into Eq. (\ref{pot}). However, as is clear from Eq.(\ref{psi}), there is no analytic way in which we could express $r$ in term of $\Psi$. 

\begin{figure}
    \centering
    \includegraphics[scale=0.3]{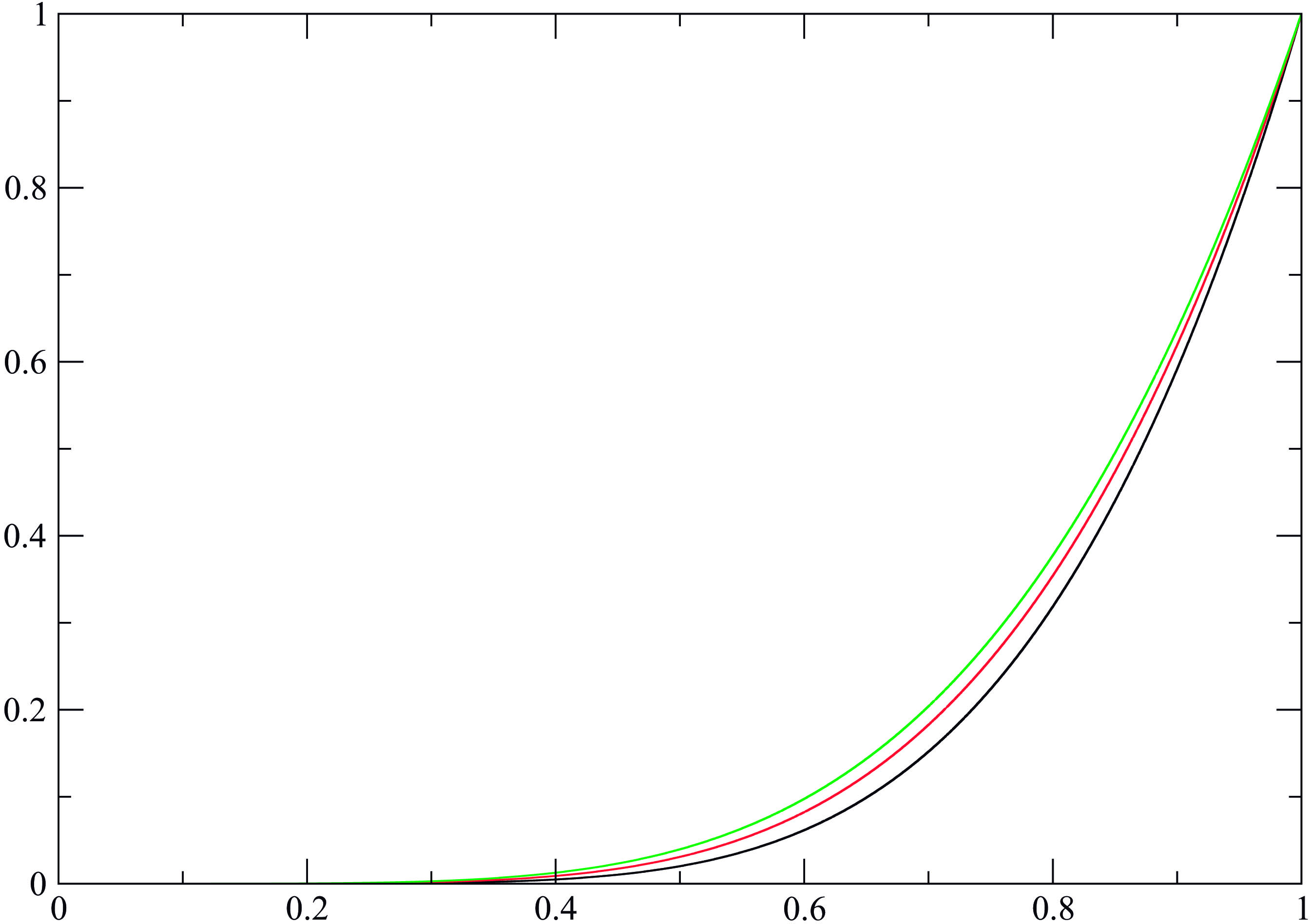}
    \caption{V/V(R) vs $\Psi^2/\Psi^2(R)$ for $M=1$, $R=5$, $\alpha C^2=0.7$, $D/\sigma=0.01$ (black curve), $D/\sigma=0.02$ (red curve) and $D/\sigma=0.03$ (green curve).}
    \label{fig:Vpf}
\end{figure}

The only possibility that we found, in which $V$ can be written in terms of $\Psi$, correspond to take limit $\sigma \rightarrow \infty $, 
\begin{equation}
    V \approx \frac{\alpha C^2}{243 k^2 M^4}\frac{\Psi^6}{K^6}\left(1-\frac{K^2}{\Psi^2}\right)^5,
\end{equation}
where
\begin{equation}
    K^2 = \frac{8\alpha C^2}{9k^2 M^2}. 
\end{equation}
This is the same result (as is expected) reported in \cite{Ovalle13}.
Now, in order to show the bahavior of $V$ as a funtion of $\Psi$ we solve Eq. (\ref{cs}) with numerical methods. In this case, will be convenient to consider $\Psi(r)=0$ if $r\rightarrow \infty$. The results for some values of the important constant are shown in figure \ref{fig:Vpf}.

\subsubsection{Second example}
As a second example we use the tidal charged black hole solution \cite{Dadhich} for the effective RSBW equations given by

\begin{eqnarray}
e^{\xi_t} & = & 1-\frac{2\,M}{r}-\frac{Q}{r^2}, \\
\mu & = & 1-\frac{2\,M}{r}-\frac{Q}{r^2}
\end{eqnarray} (the subindex $t$ is referred to ``tidal"). 

We see that 

\begin{eqnarray}
f_t^* & = & \frac{ D (r (r-2 M)-Q)}{(r (r-3 M)-2 Q)^2}, \\
\Psi'^2 & = & -\frac{2\alpha D \left(2 Q+r^2\right)}{k^2  (r (3 M-r)+2 Q)}, \nonumber \\ & \times & \frac{1}{\left(r^2 \left(\alpha D+(r-3 M)^2\right)+4 Q r (3 M-r)+4 Q^2\right)} \nonumber \\  && \label{55} \\
V_t & = & \frac{\alpha D \left(M \left(r^2-Q\right)+2 Q r\right)}{k^2  r (r (r-3 M)-2 Q)^3}.
\end{eqnarray}
Now, for $\alpha,Q<<1$ we can integrate (\ref{55}) up to first order in $\alpha$ and $Q$ to obtain
\begin{eqnarray}
\Psi^2 &\approx& \frac{16 D \alpha Q \left(-54 M^2 r+27 M^3+72 M r^2-16 r^3\right)}{243k^2  M^4 r (3 M-r)^2} \nonumber \\ &-&\frac{8 D \alpha r}{9 k^2M^2 (3 M-r)}, \label{tsp}
\end{eqnarray}
where we use, in Eq. (\ref{cs}), that
\begin{eqnarray}
\Psi_0 = \frac{\sqrt{ 8 \alpha D}  \left(16 Q-27 M^2\right)}{81 k M^3}.
\end{eqnarray}
The scalar potential up to first order in $\alpha$ and $Q$ is given by
\begin{eqnarray}
V \approx \frac{D M \alpha}{\left(k^2 r^2\right) (r-3 M)^3}+\frac{D\alpha Q \left(3 M^2-M r+2 r^2\right)}{\left(k^2 r^4\right) (r-3 M)^4}. \label{spt}
\end{eqnarray}
In order to write $V$ as in terms of $\Psi$ we need to solve Eq. (\ref{tsp}) for $r$ and introduce the result in Eq. (\ref{spt}). 

Eq. (\ref{tsp}) is cubic in $r$, that is

\begin{eqnarray}
a(\Psi^2)r^3 + b(\Psi^2)r^2 + c(\Psi^2)r + d(\Psi^2) =0, \label{re}
\end{eqnarray}
where
\begin{eqnarray*}
a(\Psi^2) &\equiv& 216 \alpha D M^2-256 \alpha D Q-243 k^2 M^4\Psi^2, \\
b(\Psi^2) &\equiv& -648 \alpha D M^3+1152 \alpha D M Q+1458 k^2 M^5 \Psi^2, \\
c(\Psi^2) &\equiv& -864 \alpha D M^2 Q-2187 k^2 M^6 \Psi^2, \\
d(\Psi^2) &\equiv& 432 \alpha D M^3 Q.
\end{eqnarray*}
From these expressions it is easy to show  (at least for $\alpha,Q<<1$) that (\ref{re}) has only one real solution for $r$ as a function of $\Psi^2$, which is 

\begin{eqnarray}
r(\Psi)= -\frac{1}{3a}\left(b+K+\frac{U_1}{c}\right),
\end{eqnarray}
where
\begin{eqnarray}
K &=& \left(\frac{U_2+(U_2^2-4U_1^3)^{1/2}}{2}\right)^{1/3}, \\
U_1 &=& b^2 - 3ac, \\
U_2 &=& 2b^3-9ab+27a^2d.
\end{eqnarray}
Thus, from
\begin{eqnarray}
V &\approx& \frac{D M \alpha}{\left(k^2 r(\Psi)^2\right) (r(\Psi)-3 M)^3} \nonumber \\&+& \frac{D\alpha Q \left(3 M^2-M r(\Psi) + 2 r(\Psi)^2\right)}{\left(k^2 r(\Psi)^4\right) (r(\Psi)-3 M)^4},
\end{eqnarray}
it is clear that $V$ is a function of even powers of $\Psi$ only. 

\begin{figure}
    \centering
    \includegraphics[scale=0.3]{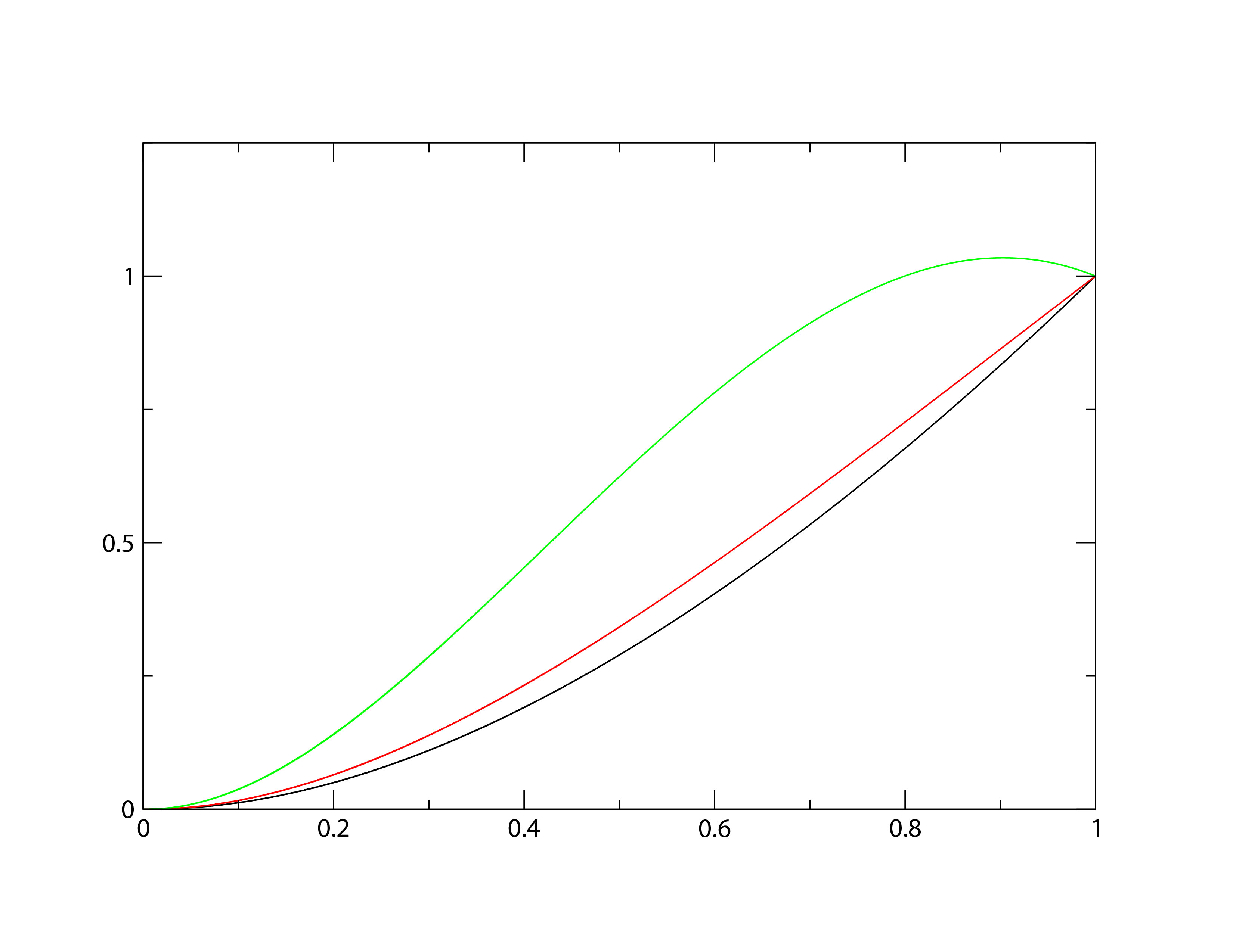}
    \caption{$V/V(R)$ vs $\Psi^2/\Psi^2(R)$ for $R=5$,$M=1$, $\alpha D=0.7$, $Q=2$ (red curve), $Q=1$ (black curve), $Q=4$ (green curve).}
    \label{fig:Vp}
\end{figure}

The scalar curvature is given by

\begin{equation}
    R=\frac{2\alpha D(2Q^2-5Qr^2+6MQr-r^4)}{r^2(r^2-3rM-2Q)^3},
\end{equation}
where we can see see that there are two singularities at $r_{\pm}=(3M\pm\sqrt{9M^2+8Q})/2$. But only $r_{+}$ is a naked singularity. However, as before, this is not a problem if $r$ is always bigger that $r_{+}$.    
We can also use numerical methods to integrate directly Eq. (\ref{55}). In figure \ref{fig:Vp} we present the results (assuming that the scalar field $\Psi$ is zero when $r\rightarrow \infty$) for some values of the important constants.

\subsubsection{Third example}
In a recent paper we found a series of external solutions for the BWRS scenario taking the tidal charged black hole solution as the seed. 

Now let us denote as $\{\xi_t,\mu_i\}$ ($i=1,2,3$; the subindex $t$ is referred to ``tidal") the metric components of the solutions obtained in \cite{Leon}, where 

\begin{eqnarray}
\mu_1 &=&
e^{\xi_t}
\left[1
+
{\beta}
\left(\frac{r}{a}\right)^2\,e^{\frac{4\,Q}{M\,r}}\left(1-\frac{M}{r}\right)^{2+\frac{4\,Q}{M^2}}
\right]
\ , \\
\mu_2
& = &
e^{\xi_t}
\left(1+\frac{B e^{\frac{3MArcTan\left[\frac{3M-4r}{\sqrt{-9M^2-8Q}}\right]}{\sqrt{-9M^2-8Q}}}}{\sqrt{r\,(2\,r-3\,M)-Q}}\right)
\ , \\
\mu_3
&=&
e^{\xi_t}
\left[1+d\,\left(1-\frac{M}{r}\right)^{\frac{2\,Q}{M^2}}\,e^{\frac{2\,Q}{M\,r}}\right]
\ .
\end{eqnarray}
Then, the coupling of these solutions with a KG scalar field are given by 

\begin{eqnarray}
\tilde{\mu}_i &=& \mu_r + \alpha f^*_t, \\
\Psi_i'^2 & = & \frac{2\alpha C^2(r^2+2Q)e^{\xi_t}}{k^2r(r^2-3Mr-2Q)^3\mu_r}, \label{fp} \\
V_i = V_t &=& \frac{\alpha D \left(M \left(r^2-Q\right)+2 Q r\right)}{4 \pi  r (r (r-3 M)-2 Q)^3}.
\end{eqnarray}
Then, as in the seed solution (tidal charge black hole), the coupling with a KG scalar field through MGD-decoupling introduce a naked singularity in new solutions if the internal matter distribution is less than $r_+$. Since the form of $\mu_r$ is much more complicated that in the first and second examples, we could not find any analytical solution to Eq. (\ref{fp}). 

\begin{figure}
    \centering
    \includegraphics[scale=0.3]{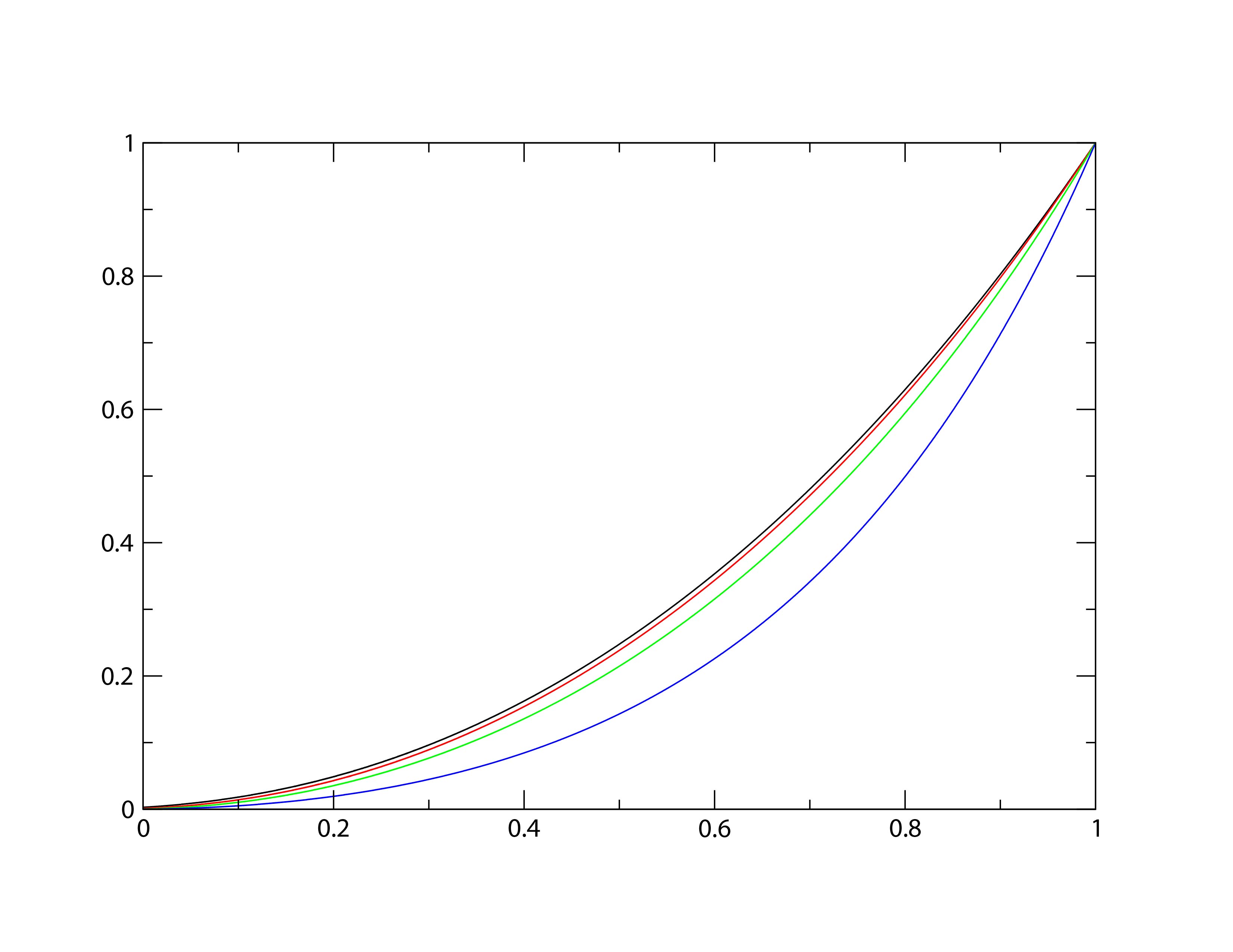}
    \caption{$V_1/V_1(R)$ vs $\Psi_1^2/\Psi_1^2(R)$ for $R=5$,$M=1$, $\alpha D=0.7$, $\beta/a^2=0.01$ and $Q=1$ (black curve), $Q=2$ (red curve), $Q=3$ (green curve), $Q=4$ (blue curve).}
    \label{fig:V1p}
\end{figure}

\begin{figure}
    \centering
    \includegraphics[scale=0.3]{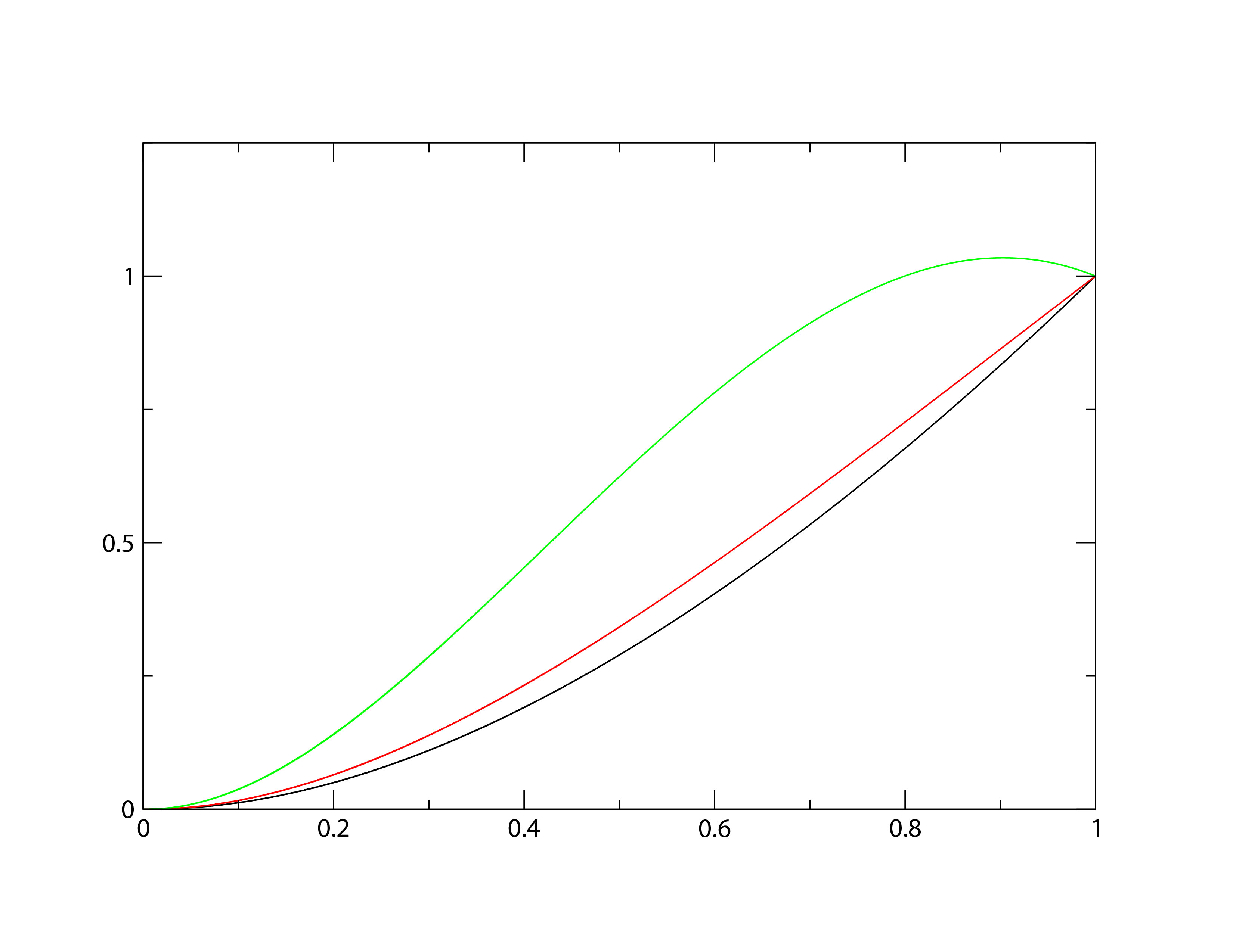}
    \caption{$V_2/V_2(R)$ vs $\Psi_2^2/\Psi_2^2(R)$ for $R=5$,$M=1$, $\alpha D=0.7$, $B=0.01$ and $Q=-1.5$ (black curve), $Q=-2$ (red curve), $Q=2.5$ (green curve).}
    \label{fig:V2p}
\end{figure}

\begin{figure}
    \centering
    \includegraphics[scale=0.3]{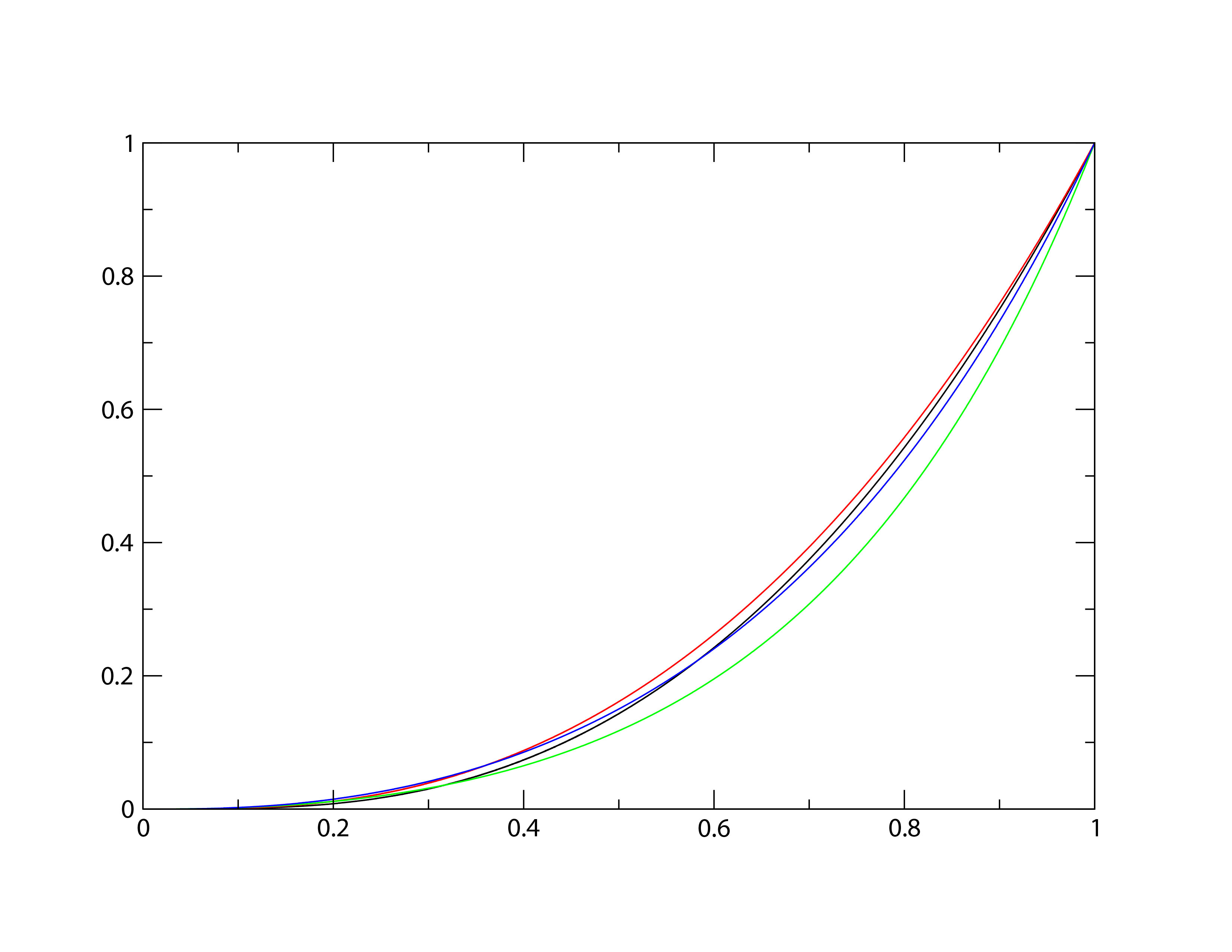}
    \caption{$V_3/V_3(R)$ vs $\Psi_3^2/\Psi_3^2(R)$ for $R=5$,$M=1$, $\alpha D=0.7$, $d=0.01$ and $Q=1.0$ (black curve), $Q=2.0$ (red curve), $Q=3.5$ (blue curve), $Q=4.0$ (green curve).}
    \label{fig:V1p}
\end{figure}

\subsection{Second approach}

Another way to couple the four dimensional effective RSBW field equations to a KG scalar field using MGD-decoupling method consists to take a known solution of the Einstein Klein Gordon field equations (EKG) and then analyze the contributions coming from extra dimensions using the mechanism developed in \cite{Leon}. 

To do that, we write the EKG field equations in the form given by

\begin{eqnarray}
k^2 \left(T^0_0 + \frac{1}{2}e^{-\lambda} \Psi'^2+V \right) & = & \frac{1}{r^2}-e^{-\lambda}\left(\frac{1}{r^2}-\frac{\lambda'}{r}\right), \\
k^2 \left(T^1_1-\frac{1}{2}e^{-\lambda} \Psi'^2+V \right) & = & \frac{1}{r^2}-e^{-\lambda}\left(\frac{1}{r^2}-\frac{\nu'}{r}\right), \\
k^2 \left(T^2_2 + \frac{1}{2}e^{-\lambda} \Psi'^2+V \right) & = & \frac{e^{-\lambda}}{4}\Big(2\nu'' + \nu'^2 -\lambda'\nu' \nonumber \\ &+& 2\frac{\nu'-\lambda'}{r}\Big).
\end{eqnarray}
By defining the effective pressures and energy density through
\begin{eqnarray}
\rho_t & = &  T^0_0 + \frac{1}{2}e^{-\lambda} \Psi'^2+V, \\
p_r  & = & -T^1_1+\frac{1}{2}e^{-\lambda} \Psi'^2-V, \\
p_t & = & -T^2_2 - \frac{1}{2}e^{-\lambda} \Psi'^2-V,
\end{eqnarray}
and taking $\alpha=1/\sigma$ it is possible to write the equations for the contributions of the bulk, obtained from the MGD method, in the form

\begin{eqnarray}
\label{ec1dch10f}
&& k^2\,\left( \frac{(\rho^2-\Delta^2)}{2}+\frac{6\mathcal{U}}{k^4}+\frac{1}{2}f^*\Psi'^2\right) =
-\frac{f^{*}}{r^2}
-\frac{f^{*'}}{r}\ ,
\\
\label{ec2dch10f}
&& k^2\,\left[\left(\frac{\rho^2}{2}+ \rho p_t + \frac{p_t^2-p^2_r}{2} +\frac{2\mathcal{U}}{k^4}\right) +\frac{4\,{\mathcal{P}}}{k^{4}}- \frac{1}{2}f^*\Psi'^2\right] \nonumber \\ && 
\hspace{3.5cm} =  f^{*}\left(\frac{1}{r^2}+\frac{\nu'}{r}\right)\ ,
\\
\label{ec3dch10f}
 &&k^2\,\left[\left( \frac{\rho ^{2}}{2} +\frac{\rho}{2}(p_r+p_t)-\frac{p_r}{2}\Delta +%
\frac{2}{k^{4}}\,\mathcal{U}\right) -\frac{2\,{\mathcal{P}}}{k^{4}} \right. \nonumber  \\ && \hspace{1cm} + \left. \frac{1}{2}f^*\Psi'^2\right]
 = \frac{1}{4}
\left[f^{*}\left(2\,\xi''+\xi'^2+2\frac{\nu'}{r}\right) \right. \nonumber \\ &&\hspace{1cm} + \left. f^{*'}\left(\nu'+\frac{2}{r}\right)\right] ,
\end{eqnarray}
with $\Delta=p_t-p_r$. From these equations, we can notice one of the big difference with the first approach. The last term of the right side in each equation is a contribution of the scalar field to the effective RSBW equations. The equilibrium equation for this case is given by 

\begin{eqnarray}
\label{conset2ch10xy}
\mathcal{U}' &+&2\mathcal{P}'+\nu'(2\mathcal{U}+\mathcal{P})+\frac{6\mathcal{P}}{r}-\frac{k^4}{2}(\rho'+p_t')(\rho+p_t) \nonumber \\ &-&\frac{k^4}{2r}(\rho+p_t)\Delta -\frac{k^4 \nu'}{4}(\rho^2+p_rp_t+\rho(p_r+p_t)) \nonumber \\ &=& -f^*\Psi'\left[\Psi''+\frac{f*'}{2f^*}\Psi'-\frac{dV}{d\Psi}+\left(\frac{\nu'}{2}+\frac{2}{r}\right)\Psi'\right].
\end{eqnarray}

Now, from Eqs. (\ref{ec1dch10f})-(\ref{ec3dch10f}), we get the following solution for $f^*$

\begin{equation}
f^*(r)=\frac{J}{F(r)}+ \frac{2k^2}{F(r)}\int \frac{F(r)r}{r\xi'+4}(\rho^2+\rho(2p_t+p_r)+\Delta^2)dr, \label{cc}
\end{equation}
with
\begin{eqnarray}
F(r) &=& \exp{\left(-\int \frac{2rI(r)}{r\xi'+4} dr\right)}, \label{fit} \\
I(r) &=& \xi''+\frac{(\xi')^2}{2}+\frac{2\xi'}{r}+\frac{2}{r^2}.
\end{eqnarray}

To show the difference of the two approaches we can select the following solution of the EKG field equations

\begin{eqnarray}
e^\nu &=& 1-\frac{2M}{r}, \\
e^{-\lambda} & = & \left(1-\frac{2M}{r}\right)\left[1+\alpha\left(\frac{C}{r-3M}\right)^2\right], \\
\Psi'^2 &=& \frac{2\alpha C^2}{k^2 r(r-3M)(\alpha C^2 + (r-3M)^2)},\\
V & = & \frac{\alpha C^2 M}{k^2  r^2 (r-3 M)^3}, 
\end{eqnarray}
which was also obtained by using the MGD method in \cite{Ovalle13}.

Introducing these expressions in the Eqs. (\ref{cc})-(\ref{fit}) (with $T^0_0=T^1_1=T^2_2=0$) and taking for simplicity $J=0$ we get

\begin{eqnarray}
F(r) &=& \frac{r(2r-3M)}{(2M-r)}, \\
f^*(r) & = & -\frac{C^4 (2 M-r)}{162 k^2 M^5 r^2 (3 M-2 r) (r-3 M)^4} (3 M \nonumber \\ &\times& (156 M^2 r^2-225 M^3 r+108 M^4 \nonumber \\ &-& 42 M r^3+4 r^4)-4 r (r-3 M)^4 \log (r) \nonumber \\ &+& 4 r (r-3 M)^4 \log (r-3 M)),
\end{eqnarray}
which again exhibits a singularity at $r=3M$.  

Nevertheless, as before, as long as $r>3M$, this represents a well-behaved external solution. The Ricci scalar for this case is given by

\begin{eqnarray}
R &=& -\frac{2 C^2}{r(r-3M)^3} + \frac{2(r-M)(r-3M)f^*}{\sigma(r-2M)^2r^2} \nonumber \\ &-&
\frac{f^{*'}}{\sigma(r-2M)^2(r-3M)^3r^2} \times (162M^5r \nonumber \\ &-& 351M^4r^2+297M^3r^3-123M^2r^4+25Mr^5-2). \nonumber \\ 
\end{eqnarray}

Now, in this case,  it is possible to find the following expression for the potential as a function for the scalar field up to first order in $\alpha$, and to all orders in $\sigma$

\begin{equation}
    V \approx \frac{\alpha C^2}{243 k^2 M^4}\frac{\Psi^6}{K^6}\left(1-\frac{K^2}{\Psi^2}\right)^5,
\end{equation}

\section*{Conclusions}
We presented the formulation of the simplest RSBW scenario model minimally coupled to a Klein Gordon scalar field in 4D using the MGD-decoupling method. The coupling with a KG scalar field could be interpreted as restricted to the visible brane or like the effective contribution of a bulk scalar field in 5D. 

We found two different ways in which the MGD-decoupling method could be applied concerning the choice of the seed solution. The most direct possibility is to take a solution of the 4D effective RSBW equations as the seed and then, through MGD-decoupling method, to couple the system with a Klein Gordon scalar field. In this work, we used  the external solutions of the effective RSBW equations obtained in \cite{Leon} as seeds. In these cases the inclusion of the scalar field introduces a naked singularity to the final solution and therefore can not represent black hole solutions. However, these external solutions could be well defined if if can be matched with an internal matter distribution with a radius greater than the radius of the singularity (see \cite{Ovalle3,Ovalle10} for more details). On the other hand, because of the BW corrections, the exact form of the potential as a functional of $\Psi$ is not clear, except for the tidal black hole solution with $\alpha<<1$.   

The other possibility to apply the MGD-decoupling method is to start with a solution of the Einstein-Klein Gordon system (in which the contributions of the scalar could be reinterpreted as an anisotropic fluid) and use results presented in \cite{Leon} to include the BW effective contributions. For example, we used the solution presented in \cite{Ovalle13} as seed. The scalar field and the potential (as a function of $\Psi$) were easier to find, at least at lower orders in $\alpha$. However, as in the previous case, the final solution exhibited the same naked singularity of the seed solution. 

In GR, the order in which we applied the MGD method to decoupled different source is not really important and the result should be the same. However, in our case, the effective contributions of the RSBW depend on the form of the energy momentum-tensor in the visible brane. This implies that the seed solution is important and, depending on the order in which we apply the method, starting from GR, will lead us into two different results. As an example, we have the first solution of the section 4.1 and the solution obtained in section 4.2. Both were found by using the same original GR seed solution. In each case, the metric components and the solution for the potential and scalar field are different. In the first approach, the potential and the scalar have more complicated expressions, while the metric components are relatively simple. The second approach lead to a solution with a much simpler form of the potential and the scalar field, however, the spacial metric component very complicated. The difference between both solutions is clear by comparing the expressions for the Ricci scalar, which is a scalar invariant, in Eqs. (95) and (99).  

To summarize, the MGD method can be used to find solutions for the $4D$ effective RSBW field equations and the Einstein-Klein Gordon system using as seed a solution of GR (see \cite{Leon,Ovalle13}). Then, as is shown in figure \ref{Diagrama}, we realized the study of the RSBW coupled to a Klein Gordon scalar field through MGD-decoupling starting with a GR solution. Nevertheless, the order in which the MGD method was applied to decouple each source showed to be fundamental.

\section{Acknowledgements}

P.L wants to say thanks for the financial support received for the financial help received by the Project ANT1956 of the Universidad de Antofagasta and CONICYT PFCHA / DOCTORADO BECAS CHILE/2019 - 21190517. P.L is grateful Semillero de Investigaci\'on SEM 18-02 from Universidad de Antofagasta and the Network NT8 of the ICTP.

\end{document}